\documentclass{desyproc}

\newcommand{\lla}{\langle\langle}
\newcommand{\rra}{\rangle\rangle}

\newcommand{\trpl}[1]{\vec{#1}_\perp}

\newcommand{\im}{\text{Im}}

\begin{document}
\title{Wilson-loop correlators on the lattice and 
  asymptotic
behaviour of hadronic total cross sections}
\author{{\slshape \underline{Matteo Giordano}$^1$, Enrico Meggiolaro$^2$,
    Niccol\`o Moretti$^3$}\\[1ex]
$^1$Institute for Nuclear Research (ATOMKI), Bem t\'er 18/c, H-4026
Debrecen, Hungary\\ 
$^2$Dipartimento di Fisica, Universit\`a di Pisa, and INFN, Sezione di
Pisa, Largo Pontecorvo 3, I-56127 Pisa, Italy\\
$^3$Institut f\"ur Theoretische Physik, Universit\"at Zurich, 8057
Z\"urich, Switzerland }

\contribID{22}


\acronym{EDS'13} 

\maketitle

\begin{abstract}
We show how universal, Froissart-like hadronic total cross sections
can be obtained in QCD in the functional-integral approach to soft 
high-energy scattering, and we discuss indications of this behaviour
obtained from lattice simulations.
\end{abstract}

\section{Introduction}

The recent measurements of hadronic total cross sections at the
LHC~\cite{TOTEM} have revived the interest in trying to understand
their behaviour at large energy from the theoretical point of
view. Experimental data support the ``Froissart-like'' rising
behaviour $\sigma_{\rm tot}^{{}_{(hh)}}(s) \sim B\log^2 s$ at large energy,
with a {\it universal} prefactor $B$, i.e., independent of the type of
hadrons involved in the scattering process~\cite{Blogs}. This
behaviour is consistent with (and named after) the well-known
Froissart-\L ukaszuk-Martin bound, which states that $\sigma_{\rm
  tot}^{(hh)}(s) \lesssim (\pi/m_\pi^2)\log^2 (s/s_0)$ for
$s\to\infty$, where $m_\pi$ is the pion mass and $s_0$ is an
unspecified scale. In principle, it should be possible to predict the
``Froissart-like'' behaviour and its universality from QCD, which we
believe to be the microscopic theory of strong interactions, but a
satisfactory derivation is still lacking.

The main reason why it is so difficult to obtain predictions for
$\sigma_{\rm tot}^{{}_{(hh)}}$ from first principles is that this requires
a better understanding of the nonperturbative (NP) dynamics of
QCD, which is known to be a hard task. Indeed, total cross sections
are part of the more general problem of soft high-energy scattering,
characterised by a large total center-of-mass energy squared $s$ and
a small transferred momentum, $|t|\lesssim 1 {\rm GeV}^2\ll s$. In this
energy regime, perturbation theory is not fully reliable, and one has
to attack the problem with NP methods. In this context, a
functional-integral approach in the framework of QCD has been proposed
in~\cite{Nachtmann91} and further developed in~\cite{DFK}, which we
now briefly recall, focussing for simplicity on the case of the
elastic scattering of two mesons of equal mass $m$. 

The elastic
meson-meson scattering amplitude ${\cal M}_{(hh)}$ is reconstructed
from the scattering amplitude ${\cal M}_{(dd)}$ of two dipoles of
fixed transverse sizes $\vec r_{1,2\perp}$, with fixed longitudinal
momentum fractions $f_{1,2}$ of the quarks, after folding with
appropriate squared wave functions $\rho_{1,2}=|\psi_{1,2}|^2$
describing the interacting hadrons~\cite{DFK},
\begin{align}
{\cal M}_{(hh)}(s,t) &=  \textstyle\int
d^2\nu\rho_1(\nu_1)\rho_2(\nu_2){\cal M}_{(dd)} (s,t;\nu_1,\nu_2)
\equiv \lla {\cal M}_{(dd)} (s,t;\nu_1,\nu_2) \rra ,
 \label{scatt-hadron}
\end{align}
where $\nu_i\!=\!(\vec{r}_{i\perp},f_i)$ denotes collectively the
dipole variables, $d^2\nu=d\nu_1d\nu_2$, $\int\!d\nu_i  = \int
d^2\vec{r}_{i\perp}\!\int_0^1 df_i$, and $\int d\nu_i\rho_i(\nu_i)=1$. 
In turn, the dipole-dipole ({\it dd\/}) scattering amplitude in
impact-parameter space is given by the (properly normalised)
correlation function (CF) of two Wilson loops (WL) in the fundamental
representation, running along the classical paths described by the 
quark and antiquark in each dipole,
thus forming a hyperbolic angle $\chi \simeq \log(s/m^2)$ in the
longitudinal plane, and properly closed by straight-line ``links'' in the
transverse plane in order to ensure gauge invariance. Eventually, one
has to take loops of infinite longitudinal extension.
The relevant Minkowskian CF ${\cal
  C}_M(\chi;\vec{z}_\perp;\nu_1,\nu_2)$, where $\vec{z}_\perp$ is
the {\it impact parameter}, i.e., the transverse 
distance between the dipoles, can be reconstructed by
means of {\it analytic continuation} from the Euclidean CF of two
Euclidean WL, ${\cal C}_E(\theta;\vec{z}_\perp;\nu_1,\nu_2) \!\equiv\! 
\lim_{T\to\infty}\langle {\cal W}^{{}_{(T)}}_1 {\cal W}^{{}_{(T)}}_2\rangle/(\langle
{\cal W}^{{}_{(T)}}_1 \rangle \langle {\cal W}^{{}_{(T)}}_2 \rangle )
- 1$, where $\langle\ldots\rangle$ is the average in the sense of the
Euclidean QCD functional
integral~\cite{Meggiolaro97,Meggiolaro05,GM2009}. 
The relevant
Euclidean WL form  
an angle $\theta$ in the longitudinal plane, while having the very
same configuration in the transverse plane as in Minkowski
space.\footnote{More precisely, the relevant paths are obtained by
  connecting the quark $[q]$-antiquark $[\bar{q}]$
straight-line paths, ${\cal C}_i : 
{X}_i^{{}_{q[\bar{q}]}}(\tau) = {z}_i + \frac{{p}_{i}}{m} \tau +
f^{{}_{q[\bar{q}]}}_i {r}_{i}$, $i =1,2$,
with $\tau\in [-T,T]$, by means of straight-line paths in the
transverse plane at $\tau\!=\!\pm T$. Here
${p}_{1,2}={m}(\pm\sin\frac{\theta}{2}, \vec{0}_{\perp},
\cos\frac{\theta}{2})$, ${r}_{i}=(0,\vec{r}_{i\perp},0)$, ${z}_i=
\delta_{i1}(0,\vec{z}_{\perp},0)$, and
 $f_i^{{}_{q}} \equiv 1-f_i$,
$f_i^{{}_{\bar{q}}} \equiv -f_i$. } 
The {\it dd}
scattering amplitude is then obtained from ${\cal C}_E(\theta;\ldots)$
[with $\theta\in(0,\pi)$] by means of analytic continuation as ($t =
-|\vec{q}_\perp|^2$) 
\begin{align}
 \label{scatt-loop}
 &{\cal M}_{(dd)} (s,t;\nu_1,\nu_2) 
  = -i\,2s \textstyle\int d^2 \vec{z}_\perp
  e^{i \vec{q}_\perp \cdot \vec{z}_\perp}
  {\cal C}_E(\theta\to -i\chi ; \vec{z}_\perp;\nu_1,\nu_2)\, .
\end{align}

\section{Lattice results and total cross sections}

In Euclidean space one can compute ${\cal C}_E$ exploiting the
available NP techniques, including the Stochastic Vacuum Model
(SVM)~\cite{LLCM2}, the Instanton Liquid Model (ILM)~\cite{ILM,GM2010}, the
AdS/CFT correspondence for planar ${\cal N}=4$ SYM~\cite{JP}, and in
particular {\it Lattice Gauge Theory} (LGT), which allows to obtain
by means of Monte Carlo simulations the true QCD prediction for the CF
${\cal C}_E$ (within the errors). In Refs.~\cite{GM2008,GM2010},
${\cal C}_E$ was computed numerically in {\it quenched} QCD on a
$16^4$ hypercubic lattice at lattice spacing $a\simeq 0.1\,{\rm fm}$,
for loops of transverse size $a$ and with $f_{1,2}=1/2$ (which causes
no loss of generality~\cite{GM2010}), at distances
$|\vec{z}_{\perp}|/a=0,1,2$, for several angles $\theta$ and different 
configurations in the transverse plane. These included the one
relevant to meson-meson scattering, where the
orientation of dipoles is averaged over (``{\it ave}''). The
comparison of the numerical results with the analytic results obtained
in QCD-related models (SVM and ILM) showed a poor agreement, both
quantitatively (comparing with the numerical predictions of the
models) and qualitatively (fitting the data with the model
functions)~\cite{GM2008,GM2010}. Moreover, these models 
do
not lead to a ``Froissart-like'' asymptotic behaviour of $\sigma_{\rm
  tot}^{{}_{(hh)}}$: SVM and ILM 
lead to constant $\sigma_{\rm tot}^{{}_{(hh)}}$, while
the AdS/CFT expression leads to power-like $\sigma_{\rm
  tot}^{{}_{(hh)}}$~\cite{GP2010}.


In~\cite{GMM} we introduced and partially justified a class of
parameterisations of the lattice data 
that lead to Froissart-like and universal $\sigma_{\rm tot}^{{}_{(hh)}}$,
which allow to improve the best fits. These
parameterisations are of the general form $\mathcal{C}_E=\exp\{K_E\}-1$
(with $K_E$ real), with $K_E$ decaying exponentially at large
$|\trpl{z}|$, and such that after analytic continuation $K_E(\theta\to
-i\chi) \to i\, \beta(\nu_1,\nu_2)\,e^{\eta(\chi)}\,e^{-\mu |\trpl{z}|}$ at large
$\chi$ and large $|\trpl{z}|$, with $\im\, \beta \ge 0$, and $\eta$ a
real function such that $\eta\to\infty$ as $\chi\to\infty$. The exponential
form of the correlator is rather well justified: it is satisfied at
large $N_c$, where $\mathcal{C}_E\sim\mathcal{O}(1/N_c^2)$; all the
known analytical models satisfy it; the lattice data of
Refs.~\cite{GM2008,GM2010} confirm it. The exponential decay of
$K_E\sim e^{-\mu   |\trpl{z}|}$ at large impact-parameter 
is natural in a {\it confining} theory like QCD, with the relevant
mass scale $\mu$ being related to the masses of particles (including,
possibly, also glueballs) exchanged between the two WL. Finally, the
request $\im\, \beta \ge 0$ corresponds to a 
stronger version of the {\it unitarity constraint} on the impact
parameter amplitude $A(s,|\trpl{z}|)\equiv
\lla\mathcal{C}_M(\chi;\trpl{z};\nu_1,\nu_2)\rra$. 
It is known
that $|A+1|\leq 1$; 
as this
constraint has to be satisfied for all physical choices of
$\rho_{1,2}$ in Eq.~\eqref{scatt-hadron}, it is natural to assume the
strongest constraint $|\mathcal{C}_M+1|\le 1$ $\forall \trpl{z},
\nu_1,\nu_2$ at large $\chi$. 
In~\cite{GMM} we showed that our parameterisations lead to 
$\sigma^{{}_{(hh)}}_{\text{tot}} \sim
{2\pi}\textstyle\eta^2/{\mu^{2}}$ 
at large $\chi$. Taking 
$e^\eta= \chi^p e^{n\chi}$, one obtains the \textit{universal} result
$\sigma^{{}_{(hh)}}_{\text{tot}} \sim B \log^2 s$, with $B = 2\pi
n^2/\mu^2$ independently of the mesons involved in the process. 

\begin{wraptable}{r}{0.34\textwidth}
\centerline{\begin{tabular}{|l|c|c|}
\hline
   $i$    & $\mu$ (GeV)  & $B=\frac{2\pi}{\mu^2}$ (mb) \\
\hline
1 & $4.64(2.38)$  & $0.113^{+0.364}_{-0.037}$ \\
2 & $3.79(1.46)$  & $0.170^{+0.277}_{-0.081}$ \\
3 & $3.18(98)$    & $0.245^{+0.263}_{-0.100}$ \\
\hline
\end{tabular}}
\caption{Mass-scale $\mu$ and
  the coefficient $B$
obtained with our parameterisations $\mathcal{C}^{ave}_i$.}  
\label{tab:lambdavac}
\end{wraptable}
The analysis of the lattice data was performed using
$\mathcal{C}^{ave}$, which is closer to the physical amplitude (the
analysis above can be repeated for $\mathcal{C}^{ave}$ without altering
any conclusion). Our best parameterisations 
are 
{$\mathcal{C}^{ave}_i=\exp
\{K_E^{{}_{(i)}}\}-1$}, $i=1,2,3$, with 
$K_E^{{}_{(1)}}=\frac{K_1}{\sin\theta}+K_2 \cot^2\theta + K_3
\cos\theta\cot\theta$ and $K_E^{{}_{(2)}} = \frac{K_1}{\sin\theta} + K_2
(\frac{\pi}{2}-\theta) \cot\theta + K_3 \cos\theta\cot\theta$, which
are essentially two proper modifications of the AdS/CFT result (taking
into account that $\mathcal{C}^{ave}$ is symmetric under
crossing~\cite{GM2006}), and $K_E^{{}_{(3)}} = \frac{K_1}{\sin\theta}
+ K_2(\frac{\pi}{2}-\theta)^3 \cos\theta$. In the three cases, the
unitarity condition is satisfied if $K_2\geq0$: this is actually the
case for our best fits (within the errors). 
The value of $B=2\pi/\mu^2$, obtained through a fit
of the coefficient of the leading term with an exponential function,
is found to
be compatible with the experimental result (within the large errors)
in all the three cases (see Table~\ref{tab:lambdavac}). However, this
must be taken only as an estimate, as our lattice data are {\it
  quenched}, and available only for rather small $|\trpl{z}|$.

\section{Acknowledgments}

MG is
supported by MTA 
under ``Lend\"ulet''
grant No.~LP2011-011.


\begin{footnotesize}

\end{footnotesize}
\end{document}